\documentclass[a4paper,12pt]{iopart}

\usepackage{iopams,epsfig,graphicx}

\newcommand{\be}{\begin{eqnarray}}
\newcommand{\ee}{\end{eqnarray}}

\def\Ro{{\mathbb R}}

\def\Ro{{\mathbb R}}
\def\qq{{\bf q}}
\def\pp{{\bf p}}

\def\upd{{\rm d}}

\begin{document}
\title{On the thermodynamics of classical micro-canonical systems}
\author{
Maarten Baeten and Jan Naudts
}
\address{
Departement Natuurkunde, Universiteit Antwerpen,
Universiteitsplein 1, 2610 Antwerpen, Belgium
}
\ead{Maarten.Baeten@student.ua.ac.be, Jan.Naudts@ua.ac.be
}

\begin{abstract}
We study the configurational probability distribution of
a mono-atomic gas with a finite number of particles $N$ in the
micro-canonical ensemble. We give two arguments why the
thermodynamic entropy of the configurational subsystem
involves R\'enyi's entropy function rather than that of Tsallis.
The first argument is that the temperature of the
configurational subsystem is equal to that of the kinetic
subsystem. The second argument is that the instability
of the pendulum, which occurs for energies close to the rotation threshold,
is correctly reproduced.
\end{abstract}

%

\section {Introduction}

The recent interest in the micro-canonical ensemble
\cite {GD90,HA94,GD01,BH05,GLTZ05,GLTZ06,BP06,CR06,NVdS06,CRT07,KM09}
is driven by the awareness that this ensemble is the cornerstone of statistical mechanics.
Of particular interest is the occurrence of thermodynamic instabilities in closed systems
and their relation with phase transitions. The latter are usually studied in the context
of the canonical ensemble.

The present paper focuses on the configurational probability distribution
of a mono-atomic gas with $N$ interacting particles within a non-quantum-mechanical
description. Recently \cite {NB09}, it was proved that this distribution
belongs to the $q$-exponential family \cite {NJ04,NJ08,NJ09}, with $q=1-2/(3N-2)$.
In the thermodynamic limit $N\rightarrow\infty$ it converges to the Boltzmann-Gibbs
distribution. This observation places the statistical physics of real gases
into the realm of Tsallis' non-extensive thermostatistics \cite {TC88,TC09}.
In the Tsallis community the belief reigns that the Tsallis entropy function
is more appropriate than that of R\'enyi, although both are maximised by the
same probability distributions. In favour of this point of view is the
Lesche stability \cite {LB82,AS02,NJ04c} of the Tsallis entropy function.
Moreover it has been proved that this entropy function is uniquely
associated with the $q$-exponential family (up to a multiplicative and an additive constant).
However, it was argued in \cite {NB09} that for the calculation of thermodynamic
quantities the R\'enyi entropy function is more appropriate. This statement is
elaborated in the present work. In addition, it is shown by means of the example
of the pendulum
that the stability of the Tsallis entropy function makes it inappropriate
to describe instabilities of closed systems.

\begin{figure}[!h!t]
	\centering
	\includegraphics[width=6cm]{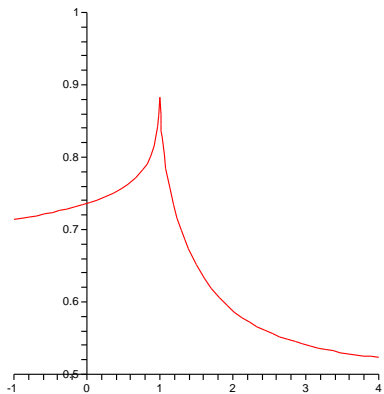}
	\caption{Density of states of the pendulum in reduced units}
	\label {figone}
\end{figure}

The pendulum is an interesting example because it has two distinct types of orbits: librational motion at
low energy and rotational motion at high energy. The density of states $\omega(U)$ can be calculated
analytically (see for instance \cite {NJ05}). It has a logarithmic singularity at the energy $U_c$,
which is the minimum value needed to allow rotational motion --- see the Figure \ref {figone}.

The thermodynamic quantity central to the micro-canonical ensemble is the entropy $S(U)$
as a function of the total energy $U$. Therefore, we start with it in the next Section.
Sections 3 to 6 discuss the configurational probability distribution and its
properties. Section 7 considers the ideal gas as a special case.
Section 8 deals with the example of the pendulum.
Finally, Section 9 draws some conclusions. The short Appendix
clarifies certain calculations.

\section {Micro-canonical entropies}

The entropy $S(U)$, which is most often used for a gas of point particles
in the classical micro-canonical ensemble, is
\be
S(U)=k_B\ln\omega(U),
\label {intro:boltzmann}
\ee
where $\omega(U)$ is the $N$-particle density of states. It is given by
\be
\omega(U)=c_N\int_{\Ro^{3N}}\upd\pp_1\cdots\upd\pp_N
\int_{\Ro^{3N}}\upd\qq_1\cdots\upd\qq_N\,
\delta(U-H(\qq,\pp)).
\ee
Here, $\qq_j$ is the position of the $j$-th particle and $\pp_j$ is the conjugated momentum,
$H(\qq,\pp)$ is the Hamiltonian. The constant $c_N$ equals $1/N!h^{3N}$. The constant $h$ has
the same dimension as Planck's constant. It is inserted for dimensional reasons.
This definition goes back to Boltzmann's idea of equal probability
of the micro-canonical states and the corresponding well-known formula
$S=k_B\ln W$, where $W$ is the number of micro-canonical states.
However, the choice (\ref {intro:boltzmann}) of the definition of entropy has some drawbacks.
For instance, for the pendulum the entropy $S(U)$ as a function of
internal energy $U$ is a piecewise convex function
instead of a concave function \cite {NJ05}.
The lack of concavity can be interpreted as a micro-canonical instability \cite {GD90,GD01}.
But there is no physical reason why the pendulum should be classified as being unstable
at all energies.

The shortcomings of Boltzmann's entropy have been noticed long ago.
A slightly different definition of entropy is \cite {SA48,PHT85}
(see also in \cite {SBJ06} the reference to the work of A. Schl\"uter )
\be
S(U)=k_B\ln\Omega(U),
\label {intro:pearson}
\ee
where $\Omega(U)$ is the integral of $\omega(U)$ and is given by
\be
\Omega(U)=c_N\int_{\Ro^{3N}}\upd\pp_1\cdots\upd\pp_N
\int_{\Ro^{3N}}\upd\qq_1\cdots\upd\qq_N\,
\Theta(U-H(\qq,\pp)).
\ee
Here, $\Theta(x)$ is Heaviside's function.
An immediate advantage of (\ref {intro:pearson}) is that
the resulting expression for the temperature $T$, 
defined by the thermodynamic formula
\be
\frac 1{T}=\frac {\upd S}{\upd U},
\label {intro:Tdef}
\ee
coincides with the experimentally used notion of temperature.
Indeed, there follows immediately
\be
k_BT=\frac {\Omega(U)}{\omega(U)}.
\label {intro:Tres}
\ee
It is well-known
that for classical mono-atomic gases the r.h.s.~of (\ref {intro:Tres}) coincides
with twice the average kinetic energy per degree of freedom.
Hence, the choice of (\ref {intro:pearson}) as the thermodynamic entropy
has the advantage that the equipartition theorem,
assigning $k_BT/2$ to each degree of
freedom, does hold for the kinetic energy also in the micro-canonical ensemble.
Quite often the average kinetic energy per degree of freedom
is experimentally accessible and provides a unique way to measure
accurately the temperature of the system.

\section {The configurational subsystem}

The micro-canonical ensemble is described by the singular probability density function
\be
f_U(\qq,\pp)=\frac 1{\omega(U)}\delta(U-H(\qq,\pp)),
\ee
where $\delta(x)$ is Dirac's delta function.
The normalization is so that
\be
1=c_N\int_{\Ro^{3N}}\upd\pp_1\cdots\upd\pp_N
\int_{\Ro^{3N}}\upd\qq_1\cdots\upd\qq_N
f_U(\qq,\pp).
\ee
For simplicity, we take only one conserved quantity into account, namely the total energy.
Its value is fixed to $U$.

In the simplest case the Hamiltonian is of the form
\be
H(\qq,\pp)=
\frac 1{2m}\sum_{j=1}^N|\pp_j|^2+{\cal V}(\qq),
\label {hamilt}
\ee
where ${\cal V}(\qq)$ is the potential energy due to interaction
among the particles and between the particles and the walls of the system.
It is then possible to integrate out the momenta.
This leads to the configurational probability distribution, which is given by
\be
f_U^{\rm conf}(\qq)
&=&\left(\frac ah\right)^{3N}\int_{\Ro^{3N}}\upd\pp_1\cdots\upd\pp_N\,f_U(\qq,\pp).
\label {fconf}
\ee
The normalization is so that
\be
1&=&\frac 1{N!a^{3N}}\int_{\Ro^{3N}}\upd\qq_1\cdots\upd\qq_N
f_U^{\rm conf}(\qq).
\ee
The constant $a$ has been introduced for dimensional reasons\footnote{
Note that the normalisation here differs from that in \cite {NB09}}.
In the limit of an infinitely large system, this configurational system is described by
a Boltzmann-Gibbs distribution. However, here we are interested in small systems
where an exact evaluation of (\ref {fconf}) is necessary. A straightforward calculation yields
\be
f_U^{\rm conf}(\qq)
&=&\frac {[U-{\cal V}(\qq)]_+^{3N/2-1}}{\epsilon^{3N/2}\omega (U)\Gamma(3N/2)},
\label {explic}
\ee
with $\epsilon=h^2/2\pi ma^2$.

\section {The variational principle}

It was shown in \cite {NB09} that the configurational probability distribution
$f_U^{\rm conf}(\qq)$ belongs to the $q$-exponential family, with $q=1-\frac 2{3N-2}$.
This implies \cite {NJ04,NJ08,NJ09} that it maximizes the expression
\be
I(f)-\theta U^{\rm conf}(f)
\label {varprin}
\ee
for some value of $\theta$, where
\be
I(f)=-\frac {1}{1-q} \frac 1{N!a^{3N}}\int_{\Ro^{3N}}\upd\qq_1\cdots\upd\qq_N\,
f(\qq)
\left[\left(f(\qq)\right)^{1-q}-1\right]
\label {tsallis}
\ee
with
\be
U^{\rm conf}(f)=\frac 1{N!a^{3N}}\int_{\Ro^{3N}}\upd\qq_1\cdots\upd\qq_N\,
f(\qq) \, {\cal V}(\qq).
\ee
This is called the {\sl variational principle}.
Note that $I(f)$ is the Tsallis entropy function \cite {TC88} up to one modification
(replacement of the parameter $q$ by $2-q$).
The parameter $\theta$ turns out to be given by
\be
\theta=\frac {2-q}{1-q}\,\frac {1}{\epsilon^{2-q}[\Gamma(3N/2)\omega(U)]^{1-q}}.
\ee

The maximisation of (\ref {varprin}) is equivalent to the minimisation of the free energy
(using $I(f)$ as the entropy function appearing in the definition of the free energy).
Replacing the Boltzmann-Gibbs-Shannon (BGS) entropy function by $I(f)$ is necessary --- the configurational
probability distribution $f_U^{\rm conf}(\qq)$ does {\sl not} maximize the BGS entropy function 
as a consequence of finite size effects.

Note that in \cite {NJ08} the definition (\ref {tsallis}) of the entropy function contains
an extra factor $1/(2-q)$ to fix its normalisation and to make it unique within a class
of properly normalised entropy functions. This normalisation factor is not wanted in the present paper
because it becomes negative when we use $q=3$ in the example.

\section {R\'enyi's entropy function}

It is tempting to identify the parameter $\theta$ of the previous Section
with the inverse temperature $\beta=1/k_{\rm B}T$
and to interpret (\ref {varprin}) as the maximisation of the entropy
function $I(f)$ under the constraint that the average energy $U^{\rm conf}(f)$
equals the given value $U^{\rm conf}$. However, in \cite {NB09} an example
was given showing that this identification of $\theta$ with $\beta$ cannot be correct in general.
It was noted that
replacing the Tsallis entropy function by that of R\'enyi gives a more satisfactory result.
This argument is now repeated in a more general setting.

In the present context, R\'enyi's entropy function of order $\alpha$ is defined by
\be
I^\alpha(f)=\frac 1{1-\alpha}\ln\left[
\frac 1{N!a^{3N}}\int_{\Ro^{3N}}\upd\qq_1\cdots\upd\qq_N\,f(\qq)^{\alpha}
\right].
\label {renyi}
\ee
Let $\alpha=2-q$. Then (\ref {renyi}) is linked to (\ref {tsallis}) by
\be
I^\alpha(f)=\xi(I(f))
\ee
with
\be
\xi(u)=\frac 1{q-1}\ln [1+(q-1)u].
\ee
Note that
\be
\frac {\upd\xi}{\upd u}=\frac 1{1+(q-1)u}.
\ee
This derivative is strictly positive on the domain of definition of $\xi(u)$.
Hence, $\xi(u)$ is a monotonically increasing function.
Therefore, the density $f(\qq)$ is a maximizer
of $I(f)$ if and only if it maximizes $I^\alpha(f)$. This means that from the point of view
of the maximum entropy principle it does not make any difference whether one uses the R\'enyi
entropy function or that of Tsallis. However, for the variational principle discussed in the
previous Section, and for the definition of the temperature $T$ via the thermodynamic
formula (\ref {intro:Tdef})
the function $\xi(u)$ makes a difference. In the example of the pendulum, discussed further on,
the variational principle is not satisfied when using R\'enyi's entropy function, while it is
satisfied when using $I(f)$.
Also, the derivation which follows below shows that,
when R\'enyi's entropy function is used,
the temperature of the configurational
subsystem equals the temperature $T$ of the kinetic subsystem.

\section {Configurational thermodynamics}
\label {sect:conftherm}

Let us now calculate the value of R\'enyi's entropy function for the configurational
probability distribution (\ref {fconf}). One has
\be
I^\alpha(f_U^{\rm conf})
=\frac 1{q-1}\ln\left[\frac 1{N!a^{3N}}
\int_{\Ro^{3N}}\upd\qq_1\cdots\upd\qq_N\,
\left(f_U^{\rm conf}(\qq)\right)^{2-q}
\right].
\label {temp}
\ee
Use now that (see (\ref {explic}))
\be
\left(f_U^{\rm conf}(\qq)\right)^{1-q}
&=&\frac {U-{\cal V}(\qq)}{\epsilon[\Gamma(3N/2)\epsilon\omega(U)]^{1-q}}.
\ee
Then (\ref {temp}) simplifies to
\be
I^\alpha(f_U^{\rm conf})
&=&\ln \Gamma\left(\frac {3N}2\right)\epsilon\omega(U)
-\left(\frac {3N}2-1\right)\ln \frac {U^{\rm kin}}{\epsilon}.
\label {confent}
\label {claim}
\ee

The claim is now that (\ref {confent}), when multiplied with $k_{\rm B}$,  is the thermodynamic
entropy $S^{\rm conf}(U)$ of the configurational subsystem.
Note that $\epsilon=h^2/2\pi ma^2$ is an arbitrary unit of energy.
Note also that, using Stirling's approximation and $U^{\rm kin}=3N\Omega(U)/2\omega(U)$,
(\ref {claim}) simplifies to
\be
\frac 1{k_{\rm B}}S^{\rm conf}(U)
&\simeq&\ln\epsilon\omega(U)-\left(\frac {3N}2-1\right)\ln\frac {\Omega(U)}{\epsilon\omega(U)}\cr
& &+\left(\frac {3N}2-1\right)\ln \left(\frac {3N}2-1\right)
-\frac {3N}2+\frac 12\ln 3\pi N.
\ee

To support our claim, let us calculate its prediction for the temperature
of the configurational subsystem.
One finds
\be
\frac 1{T^{\rm conf}}
&\equiv&\frac {{\upd S^{\rm conf}}} {\upd U^{\rm conf}}
=k_{\rm B}\left[\frac {\omega'}{\omega}-\left(\frac {3N}2-1\right)\frac 1{U^{\rm kin}}\right]
\frac {\upd U}{\upd U^{\rm conf}}\cr
& &
+k_{\rm B}\left(\frac {3N}2-1\right)\frac 1{U^{\rm kin}}.\cr
& &
\ee
Using (see (26) of \cite {NB09})
\be
\frac {\upd U^{\rm conf}} {\upd U}=1-\frac {3N}2+\frac {\omega'}{\omega} U^{\rm kin}
\ee
this becomes
\be
\frac 1{T^{\rm conf}}
=k_{\rm B}\frac {3N}{2 U^{\rm kin}}=k_{\rm B}\frac \omega\Omega=\frac 1T.
\ee
This shows that the configurational temperature, calculated starting from
R\'enyi's entropy function, coincides with the temperature $T$ defined
by means of the modified Boltzmann entropy (\ref {intro:pearson})
and with the temperature of the kinetic subsystem. One concludes that
the natural choice of entropy function for the configurational subsystem is
R\'enyi's with $\alpha=2-q=3N/(3N-2)$.

Finally, let us define a configurational heat capacity by
\be
C^{\rm conf}=\frac {\upd U^{\rm conf}}{\upd T}=T\frac {\upd S^{\rm conf}}{\upd T}.
\ee
Then one has in a trivial way
\be
C=\frac {\upd U}{\upd T}=\frac {\upd\,}{\upd T}\left[\frac 32Nk_{\rm B}T+U^{\rm conf}\right]=\frac 32Nk_{\rm B}+C^{\rm conf}.
\ee

\section {The ideal gas}

Let us verify that the expression (\ref {claim}) makes sense even for an ideal gas.
In this case the density of states is
\be
\omega(U)=\frac {c_N}{\Gamma(3N/2)}V^N\left(2\pi m\right)^{3N/2}U^{3N/2-1}.
\ee
Evaluation of (\ref {claim}) with $U^{\rm kin}=U$ then gives
\be
S^{\rm conf}=k_{\rm B}N\ln\frac V{Na^3}+k_{\rm B}\ln\frac {N^N}{N!}
\simeq k_{\rm B}N\ln\frac {eV}{Na^3}.
\label {igCE}
\ee
The configurational entropy of an ideal gas does not depend on the total energy
or on the mass of the particles, as expected.
The total entropy is
\be
S&=&k_{\rm B}\ln\Omega(U)\cr
&=&
k_{\rm B}N\ln V
+\frac 32Nk_{\rm B}\ln 2\pi mU\cr
& &+k_{\rm B}\ln c_N
-k_{\rm B}\ln\Gamma(3N/2+1).
\label {igentropy}
\ee
Using Stirling's approximation,
(\ref {igentropy}) simplifies to
\be
S\simeq k_{\rm B}N\left(\ln \frac {eV}{Na^3}
+\frac 32\ln 2\pi\frac U{N\epsilon}+\mbox{ constant}\right).
\label {igST}
\ee
This expression coincides with the Sackur-Tetrode equation \cite {McQ00}
for an appropriate choice of the constant term.
The first term of (\ref {igST}) is the configurational entropy contribution (\ref {igCE}),
the second term is the kinetic energy contribution.

\section {The pendulum}

Let us now consider the example of the pendulum.
The Hamiltonian reads
\be
H=\frac 1{2I}p^2-\kappa^2I\cos(\phi).
\ee
For low energy $-\kappa^2I<U<\kappa^2I$ the motion is oscillatory.
At large energy $U>\kappa^2I$ it rotates
in one of the two possible directions. The density of states $\omega(U)$ can be written as
\be
\omega(U)
&=&\frac {\upd\,}{\upd U}\frac {2\sqrt{2I}}h\int\upd\phi\,\sqrt{U+I\kappa^2\cos\phi}\cr
&=&\frac {4\pi\sqrt 2}{h\kappa}\,\omega_0(U/\kappa^2I).
\ee
with $\omega_0(u)$ given by
\be
\omega_0(u)
&=&\frac 1{2\pi}\int_0^1\upd x\,\frac 1{\sqrt x}\frac 1{\sqrt {1-x}}\frac 1{\sqrt{1-u+(1+u)x}}
\qquad \mbox{ if }\quad -1<u<1,\cr
\omega_0(u)
&=&\frac 1{2\pi}\int_{-1}^1\upd x\,\frac 1{\sqrt{1-x^2}}\frac 1{\sqrt{x+u}}
\qquad \mbox{ if }\quad 1<u.
\label {pend:dos}
\ee
See the Figure \ref {figone}.
Note that the integrals appearing in (\ref {pend:dos}) are complete elliptic integrals of the first kind.
For simplicity we chose now units in which $\kappa^2I=1$ holds. We also fix $h=4/\kappa$.

\begin{figure}[!h!t]
	\centering
	\includegraphics[width=8cm]{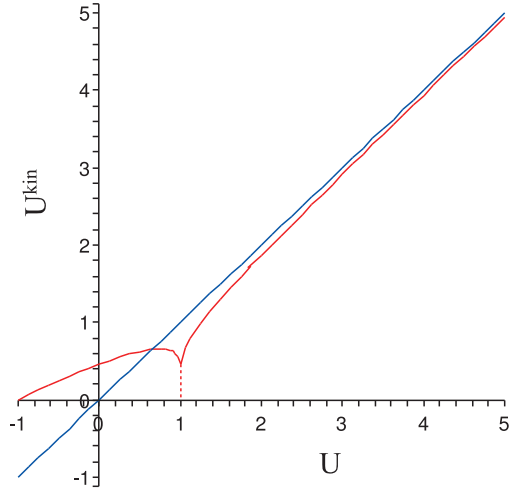}
	\caption{Kinetic energy $U^{\rm kin}$ as a function of energy $U$}
	\label {figtwo}
\end{figure}

\begin{figure}[!h!t]
	\centering
	\includegraphics[width=8cm]{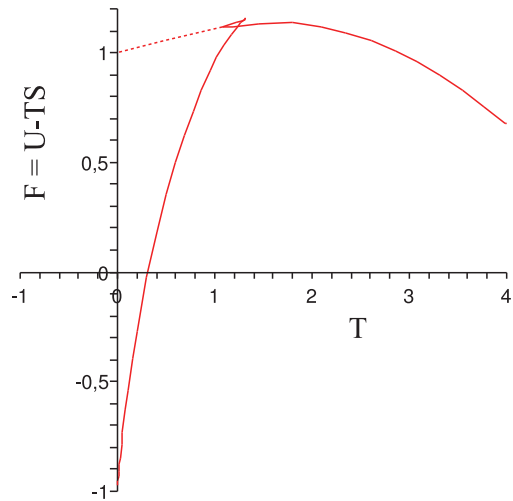}
	\caption{Free energy of the pendulum}
	\label {figthree}
\end{figure}

Using the analytic expressions (\ref {pend:dos}),
and the expression (\ref {intro:Tres}),
it is straightforward to make a plot of the temperature $T$ as a function of the energy $U$.
See the Figure \ref {figtwo}. Note that it is not a strictly increasing function. Due to the divergence of
$\omega_0(u)$ at $u=1$ the temperature $T$ has to vanish at both $u=-1$ and $u=+1$.
Hence it has a maximum in between. As a consequence, the free energy $F$,
which is the Legendre transform of $U(S)$,
\be
F(T)=U-TS
\label {freeenerg}
\ee
is a multi-valued function, when calculated by substituting $U(T)$ in (\ref {freeenerg}).
See the Figure \ref{figthree}.

When a fast rotating pendulum slows down due to friction then its energy decreases slowly.
The average kinetic energy, which is the temperature $\frac 12k_{\rm B}T$, tends to zero when
the threshold $U_c$ is approached. In the Figure \ref{figthree}, the continuous curve is followed.
The pendulum goes from a stable into a metastable rotational state. then it switches to
an unstable librating state, characterised by a negative heat capacity. Finally it goes through
the metastable and stable librational states. The first order phase transition cannot take place because
in a nearly closed system the pendulum cannot get rid of the latent heat. Neither can it stay
at the phase transition point because a coexistence of the two phases cannot be realised.

\section {The configurational free energy of the pendulum}

For the example of the pendulum
the number of degrees of freedom $3N$ in the expression for the non-extensivity
parameter $q$ has to be replaced by 1, so that $q=3$ results. This is an anomalous value
because $0<q<1$ has been assumed in the main part of the paper. See the Appendix
for a discussion of the modifications needed to treat this situation.

\begin{figure}[!h!t]
	\centering
	\includegraphics[width=8cm]{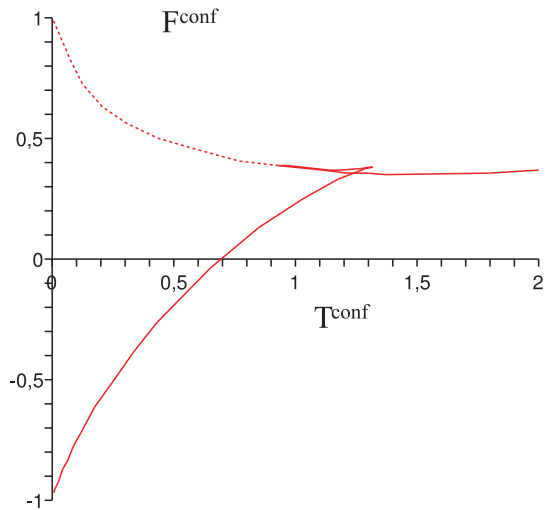}
	\caption{Configurational free energy of the pendulum}
	\label {figfour}
\end{figure}

It remains true that the configurational probability distribution
$f_U^{\rm conf}(\phi)$ maximizes the R\'enyi entropy with $\alpha=2-q=-1$
within the set of all probability distributions having the same average
potential energy $U^{\rm conf}$. Next, using 
\be
\frac 1{T^{\rm conf}}
&\equiv&\frac {{\upd S^{\rm conf}}} {\upd U^{\rm conf}}
\ee
as the definition of the temperature $T$ of the configurational subsystem,
one can plot the configurational free energy as a function of $T$.
See the Figure \ref{figfour}.

One observes the same behaviour as in the Figure \ref {figthree}.
The main difference is that in the rotational phase
the configurational free energy is a convex rather than a concave function
of the temperature $T$. This implies that the configurational
entropy $S^{\rm conf}$ is a decreasing function of $T$ and that
the heat capacity $C^{\rm conf}=T(\upd S^{\rm conf}/\upd T)$ is negative.
This is not in contradiction with the physical intuition that
the fluctuations in potential energy decrease with increasing energy $U$.
The instability of the configurational subsystem in the rotational
phase is more than compensated by the stability of the kinetic subsystem,
so that the free energy $F(T)$ of the total system is concave.

\section {Conclusions}

In a previous paper \cite {NB09} we have shown that the configurational probability distribution
$f_U^{\rm conf}(\qq)$ of a real mono-atomic gas with $N$ particles always 
belongs to the $q$-exponential family, with $q=1-\frac 2{3N-2}$.
In the same paper it was argued, based on one example, that for the
definition of the configurational temperature $T$ the entropy function of R\'enyi is better suited
than that of Tsallis. Here we show in the Section \ref {sect:conftherm}
that the same result holds for any real gas with a Hamiltonian of the usual form (\ref {hamilt}).

It is well-known that R\'enyi's entropy function and that of Tsallis are related
because each of them is a monotone function of the other. Hence, from the point of view of the
maximum entropy principle the two entropy functions are equivalent.
However, from the point of view of the variational principle (this is, the
statement that the free energy is minimal in equilibrium) the two are not
equivalent. This raises the need to distinguish between them.
The result of the Section \ref {sect:conftherm} then suggests that from
a thermodynamic point of view R\'enyi's entropy function is the preferred choice.

A further indication in the same direction comes from stability considerations.
In the literature of non-extensive thermostatistics one studies the notion of
Lesche stability \cite {LB82,AS02,NJ04c}. Tsallis' entropy function is Lesche-stable
while R\'enyi's is not. The present paper focusses on thermodynamic stability, by which
one usually understands the positivity of the heat capacity.

A well-known property of the Boltzmann-Gibbs distribution is that
it automatically leads to a positive heat capacity and that instabilities such as
phase transitions are only possible in the thermodynamic limit.
The entropy function which is maximised by the Boltzmann-Gibbs distribution is that of
Boltzmann-Gibbs-Shannon (BGS). The Boltzmann-Gibbs distribution is known in
statistics as the exponential family. Its generalisation, needed here,
is the $q$-exponential family. Both Tsallis' entropy function and that of R\'enyi
are maximised by members of the $q$-exponential family. However, only
Tsallis' entropy function shares with the BGS entropy function the property that
the heat capacity is always positive --- this has been proved in a very general context in \cite {NJ04}.
For this reason, one can say that the Tsallis' entropy function is a stable entropy function.
We have shown in the present paper with the explicit example of the pendulum
that R\'enyi's entropy function is not stable in the above sense.

The example of the pendulum was chosen because it exhibits two thermodynamic phases.
At low energy the pendulum librates around its position of minimal energy.
At high energy it rotates in one of the two possible directions.
In an intermediate energy range the time-averaged kinetic energy drops when the total
energy increases. If the kinetic energy is taken as a measure for the temperature
then the pendulum is a simple example of a system with negative heat capacity.
Hence, it is not such a surprise that, if we look for an instability, that
we find it in this system. But this also means that R\'enyi's entropy function
is able to describe the instability of the pendulum, while Tsallis' entropy function
is not suited for this task. This is again an indication that R\'enyi's entropy function
is an appropriate candidate for a statistical definition of the thermodynamic entropy
of small systems.

\appendix
\setcounter{section}{1}
\section*{Appendix}
The configurational probability distribution of the pendulum is given by
\be
f_U^{\rm conf}(\phi)=\frac {a\sqrt{I/2}}{h\omega(U)\sqrt{U+I\kappa^2\cos\phi}}.
\ee
It maximizes R\'enyi's entropy function with $\alpha=-1$. The maximal value equals
\be
S^{\rm conf}(U)
&=&\frac 12k_{\rm B}\ln\frac 1a\int\upd\phi\,\frac 1{f_U^{\rm conf}(\phi)}\cr
&=&\frac 12k_{\rm B}\ln\frac {h\omega(U)}{a^2\sqrt{I/2}}\int\upd\phi\,\sqrt{U+I\kappa^2\cos\phi}\cr
&=&\frac 12k_{\rm B}\ln\frac {h^2}{Ia^2}\omega(U)\Omega(U).
\ee
Therefore the inverse of the configurational temperature is given by
\be
\frac 1{T^{\rm conf}}
&=&\frac {\upd S^{\rm conf}}{\upd U^{\rm conf}}\cr
&=&k_{\rm B}\frac {\omega(U)^2+\omega'(U)\Omega(U)}{2\omega(U)\Omega(U)}\,
\frac {\upd U\,\,}{\upd U^{\rm conf}}.
\label {app:temp}
\ee
But note that
\be
\frac {\upd U^{\rm conf}}{\upd U}
&=&\frac {\upd \,}{\upd U}\left(U-\frac {\Omega(U)}{2\omega(U)}\right)\cr
&=&\frac {\omega'(U)\Omega(U)+\omega^2(U)}{2\omega^2(U)}.
\ee
Hence (\ref {app:temp}) becomes
\be
\frac 1{T^{\rm conf}}
&=&k_{\rm B}\frac {\omega(U)}{\Omega(U)}.
\ee
This shows that the temperature of the configurational subsystem coincides
with that of the kinetic subsystem --- see (\ref {intro:Tres}).

It is now straightforward to make the parametric plot of Figure \ref {figfour}
by plotting $T$ as a function of $U$ on the horizontal axis, and $F=U-TS$
on the vertical axis.

\vskip 1cm

\end{document}